\documentclass[11pt,epsf]{article}

\long\def\comment#1{\ifdim\overfullrule>0pt{\sf[{#1}]}\fi}

\usepackage{latexsym}
\usepackage{amsmath}
\usepackage{epsfig}

\newtheorem{theorem}{Theorem}
\newtheorem{definition}{Definition}

 \newtheorem{lemma}{Lemma}

\newtheorem{corollary}[theorem]{Corollary}

\newcommand{\qed}{\hspace*{\fill}\mbox{$\Box$}}

\begin{document} \bibliographystyle{alpha}
\def\proofend{\hfill$\Box$\medskip}
\def\Proof{\noindent{\bf Proof:\ \ }}
\def\Sketch{\noindent{\bf Sketch:\ \ }}
\def\eps{\epsilon}

\title{Graph-TSP from Steiner Cycles}

\author{Satoru Iwata\thanks{Department of Mathematical Informatics,
    University of Tokyo, Tokyo 113-8656, Japan.  Email: \tt{iwata@mist.i.u-tokyo.ac.jp}.}
  \and Alantha Newman\thanks{CNRS-Universit\'e Grenoble Alpes and
    G-SCOP, F-38000 Grenoble, France.  Supported in part by LabEx PERSYVAL-Lab (ANR--11-LABX-0025).
    Email: \tt{alantha.newman@grenoble-inp.fr}.} \and
  R. Ravi\thanks{Tepper School of Business, Carnegie Mellon
    University, USA.  Supported in part by
    NSF grants CCF1143998 and CCF1218382. Email: \tt{ravi@cmu.edu}.}}

\maketitle

\begin{abstract}
We present an approach for the traveling salesman problem with graph
metric based on Steiner
cycles.  A Steiner cycle is a cycle that is required to contain some
specified subset of vertices.  For a graph $G$, if we can find a
spanning tree $T$ and a simple cycle that contains the vertices with
odd-degree in $T$, then we show how to combine the classic ``double
spanning tree'' algorithm with Christofides' algorithm to obtain a TSP
tour of length at most $\frac{4n}{3}$.  We use this approach to show
that a graph containing a Hamiltonian path has a TSP tour of length at
most $4n/3$.  

Since a Hamiltonian path is a spanning tree with two leaves, this
motivates the question of whether or not a graph containing a spanning
tree with few leaves has a short TSP tour.  The recent techniques of
M\"omke and Svensson imply that a graph containing a depth-first-search
tree with $k$ leaves has a TSP tour of length $4n/3 + O(k)$.
Using our approach, we can show that a $2(k-1)$-vertex connected graph
that contains a spanning tree with at most $k$ leaves has a TSP
tour of length $4n/3$.  We also explore other conditions under which
our approach results in a short tour.

\end{abstract}

\section{Introduction}

We consider the well studied Traveling Salesman problem with graph
metric, also known as graph-TSP.  Throughout this paper, the input
graph $G=(V,E)$ is assumed to be an undirected, unweighted, 2-(vertex)
connected graph, and all edge lengths in the complete graph can be
obtained via the shortest path metric on the given graph.  Our goal is
to find a tour of minimium length that visits each vertex at least
once.  In this paper, we focus on a connection between graph-TSP and
that of finding {\em Steiner cycles}.

\subsection{Background}
Graph-TSP has received much attention recently.  Oveis Gharan, Saberi
and Singh were the first to improve on the approximation ratio of
$3/2$ by an infinitesmial, but constant,
factor~\cite{gharan2011randomized}.  This was quickly followed by the
breakthrough work of M\"omke and Svensson, who introduced a new
approach leading to a substantial improvement in the approximation
ratio~\cite{DBLP:conf/focs/MomkeS11}.  Subsequently, Mucha gave a
refined analysis of their approach, proving an approximation ratio of
$13/9$ for graph-TSP~\cite{mucha2012frac}.  More recently, {Seb{\H{o}}
  and Vygen presented an approximation algorithm with ratio $7/5$ for
  the problem~\cite{sebHo2012shorter}.

It is widely believed that an approximation ratio of at most $4/3$
should be efficiently computable.  The approach of M\"omke and
Svensson is based on setting up a circulation network and showing that
a low-cost circulation leads to a low cost TSP tour.  They obtained a
$4/3$-approximation for subcubic graphs, but high-degree graphs appear
to be more challenging for their framework.  Vishnoi recently gave a
randomized algorithm that finds a TSP tour very close to $n$ with high
probability for a $k$-regular graph when $k$ is sufficiently
large~\cite{vishnoi2012permanent}.  Our goal is to consider other
techniques that are applicable for graphs that are not low-degree or
regular.

\section{Steiner Cycles}

The Steiner cycle problem has been previously, but not extensively,
studied under varying definitions~\cite{cornuejols1985traveling,
DBLP:journals/eor/Gonzalez03,
  steinova2012approximability}.  For our purposes, a Steiner cycle is
defined to be a simple cycle that contains a specified subset $S \subseteq
V$ of vertices.  It may also contain any subset of vertices from the set $V
\setminus{S}$.  We use the following definition:
\begin{definition}
Given a graph $G=(V,E)$ and a subset of vertices, $S \subseteq V$, 
a {\em Steiner cycle}, $C \subset E$, is a simple cycle whose vertices
contains the set $S$.
\end{definition}
It is important to observe that in our definition of a Steiner cycle,
there are no repeated vertices, since a Steiner cycle is a simple
cycle.  We define an approximate Steiner cycle as one in which we are
allowed to repeat vertices.  For a cycle $C$, we will use $|C|$
to denote the number of unique vertices it contains.  We define
the cycle {\em length}, $\ell(C)$, to be total length of a
traversal of the cycle.  If $C$ is a simple cycle, then $|C| = \ell(C)$.
For example, in Figure \ref{fig:simp}, the non-simple cycle has eight
unique vertices and has length ten.  Our definition
of cycle length is the same as the standard definition for the
length of a TSP tour in the graph metric.
\begin{figure}[h!]
\begin{center}
\epsfig{file=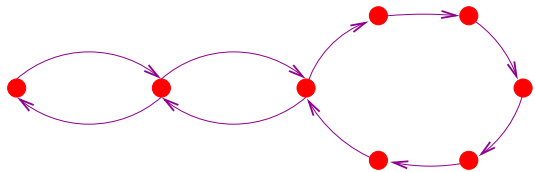, width=7cm} 
\end{center}
\caption{
In this (non-simple) cycle $C$, the number of unique vertices $|C| =
8$, but the length of the cycle $\ell(C) = 10$.}\label{fig:simp}
\end{figure}
Now we can define an approximate Steiner cycle.
\begin{definition}
Given a graph $G=(V,E)$ and a subset of vertices, $S \subseteq V$, an
{\em approximate Steiner cycle}, $C \subset E$, with relative length
$\beta \geq 1$ is a cycle whose vertices
contains the set $S$ and for which $\ell(C)/|C| \leq \beta$.
\end{definition}
In an approximate Steiner cycle, since we are allowed to repeatedly visit
vertices as we traverse the cycle, it may be the case that
the number of unique vertices will be smaller than the
length, $|C| < \ell(C)$.  Throughout this paper, whenever we refer
simply to a ``cycle'', we mean a simple cycle.

Other natural definitions of the Steiner cycle problem are concerned
with such aspects as minimizing the number of non-required (Steiner)
vertices in the cycle.  In our definition of the approximate Steiner
cycle problem, the only objective that we wish to minimize is the
ratio of the length of a cycle, $\ell(C)$, to the number of unique
vertices, $|C|$, it contains.  Thus, the measure of an optimal
solution is independent of the size of the set of required vertices.
The work that appears to be most related to the Steiner cycle problem
as we have defined it concerns the concept of {\em cyclability}: A set
of vertices $X \subseteq V$ is called {\em cyclable} if it is
contained in some cycle.  The quantity $cyc(G)$ is the maximum number
such that all subsets containing at most $cyc(G)$ vertices are
cyclable.  Note that $cyc(G) = n$ if and only $G$ is Hamiltonian.  It
seems that most of the work on cyclability has been done with the
intention of eventually using it to prove that certain graphs are
Hamiltonian or because it can be viewed as a relaxation of
Hamiltonicity.  An interesting list of theorems on cyclability can be
found in \cite{ozeki2008degree}.  Here, we explore cyclability as a
tool to obtain approximate TSP tours.

\subsection{Our Approach}

Graph-TSP can clearly be cast as a special case of the Steiner cycle
problem in which all of the vertices in $V$ are required to belong to
the Steiner cycle.  In this paper, we show that even if the required
set of vertices is possibly much smaller than the entire vertex set
$V$, an (approximation) algorithm for the Steiner cycle problem can
still be used to approximate graph-TSP.

Suppose we can find a spanning tree $T$ for the graph $G$ and a simple
cycle $C_T$ that contains all of the vertices that have an odd-degree
in the tree $T$.  When $|C_T|$ is large, we show that we can use the
folklore ``double spanning tree'' algorithm to find a short tour.
When $|C_T|$ is small, then there is a small matching on the
odd-degree vertices in $T$ and we can therefore show that Christofides
algorithm~\cite{christofides1976worst} yields a short tour.  Thus, our
algorithm, described in Section \ref{sec:main}, can be viewed as a
combination of these two standard algorithms for graph-TSP.

We are not aware of any previous work studying how to combine these
two classic algorithms for graph-TSP.  However, a similar algorithm
that combines these two algorithms was given by Guttman-Beck, Hassin,
Khuller and Raghavachari for the $s,t$-path
TSP~\cite{guttmann2000approximation}.  In their algorithm, they first
find an MST for the input graph.  If the path from $s$ to $t$ in this
MST is long, they double edges in the MST that do not belong to this
path.  If the path from $s$ to $t$ is short, they modify the input
graph by adding an edge from $s$ to $t$ with length equal to the
shortest $s,t$-path in $G$ and run Christofides on this modified graph
as in the algorithm by Hoogeveen~\cite{Hoogeveen}.  Taking the better
of these two algorithms results in a $5/3$-approximation for the
$s,t$-path problem, which does not improve on the worst-case
approximation ratio of Hoogeveen's algorithm.  Nevertheless, this
approach was used to design algorithms for special variants of the
path TSP problem~\cite{guttmann2000approximation}, and the ideas were
also eventually used to obtain improved approximation guarantees for
the $s,t$-path TSP itself~\cite{sebo2013eight}.  In our algorithm,
rather than basing the subcases on the path length from $s$ to $t$ in
an MST, we are basing the two subcases on the length of a cycle
containing the nodes with odd degree in a particular MST.

\subsection{Overview of our Results}

In Section \ref{sec:main}, we give a complete description of our
algorithm.  In Section \ref{sec:hampath}, we use this algorithm to
show that if the input graph contains a Hamiltonian path, then it has
a TSP tour of length at most $4n/3$.  Moreover, if we are given the
Hamiltonian path, then we can efficiently find such a tour.  This
theorem was first proved by Gupta using a different
approach~\cite{SwatiGms}.

One can view a Hamiltonian path as a spanning tree with two leaves.  A
natural question is how well we can approximate a TSP tour in a graph
that contains a spanning tree with few leaves.  In Section
\ref{sec:hampath2}, we show how our approach can be used to address
this question in some special cases.  In Section \ref{sec:general}, we
discuss how approximate Steiner cycles can also be used to obtain an
approximation guarantee for graph-TSP.  Finally, in Section
\ref{sec:discussion}, we consider some examples.

\begin{figure}[t!]
\begin{center}
\epsfig{file=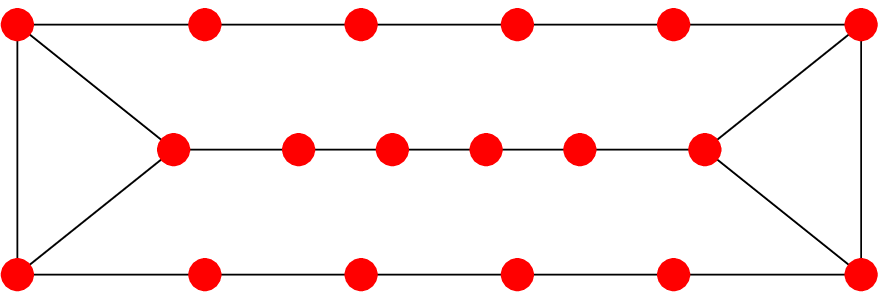, width=4cm}
\hspace{10mm}
\epsfig{file=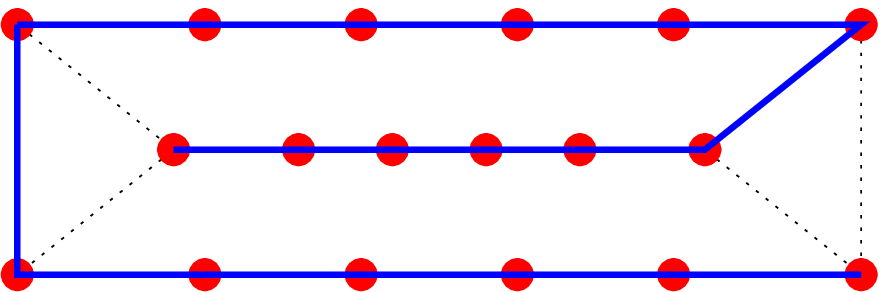, width=4cm}
\hspace{10mm}
\epsfig{file=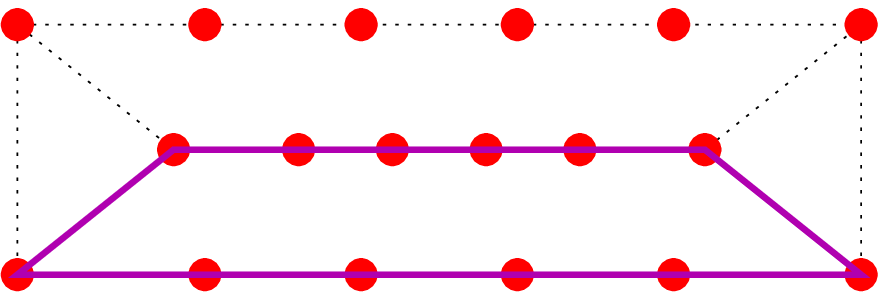, width=4cm}
\end{center}
\caption{A graph $G$ with a spanning tree $T$ (second figure, blue
  edges) and a simple cycle $C_T$ (third figure, purple edges)
  containing all of the odd-degree nodes of $T$.}
\end{figure}

\section{TSP Tours from Steiner Cycles}\label{sec:main}

Given an undirected, unweighted graph, $G=(V,E)$, with graph metric,
our goal is to find a TSP tour of minimum length.  A TSP tour must
visit each vertex at least once.  As stated previously in the
introduction, we assume that $G$ is a
2-connected graph and we define $n = |V|$.

Let $T$ be a spanning tree of $G$ and let $S_T \subset V$ be the
vertices that have odd degree in $T$.  Suppose there is a simple cycle
$C_{T}$ that contains all the vertices in $S_{T}$.  Note that the
simple cycle $C_T$ can be of arbitrary length, i.e. can contain
arbitrarily many vertices in $V\setminus{S_T}$.  

\begin{theorem}\label{cycle}
For a given graph $G$, suppose we have a minimum spanning tree $T$ and a
simple cycle $C_{T}$ that contains all vertices with odd degree in
$T$.  Then we can construct a TSP tour of $G$ with length at most
$4n/3$.
\end{theorem}

\Proof Consider the following cases.  Recall that $|C_T|$ denotes the number of
unique vertices contained in the cycle $C_T$.  Since $C_T$ is a simple
cycle, $|C_T|$ also denotes its length.
\begin{itemize}
\item[(i)] {$|C_T| > 2n/3$.  
In this case, we can contract the cycle $C_T$ to a
single vertex.  The resulting graph has at most most $n/3$ vertices.
We can then find a minimum spanning tree on this graph and double each
edge.  When we uncontract the vertex corresponding to the cycle $C_T$,
we obtain an Eulerian tour whose total length is at most $4n/3$.}

\item[(ii)] {$|C_T| \leq 2n/3$.  
In this case, since all of the vertices
of $S_{T}$ are contained in $C_T$, there is a matching of the vertices in
$S_{T}$ with length at most $n/3$.  Using this matching plus $T$, we
obtain an Eulerian tour of $G$ of length at most $4n/3$.}
\end{itemize}
\qed

We can therefore see that if $G$ has a tree $T$ and a simple cycle
$C_T$ that contains all of the vertices with odd degree in $T$,
then $G$ has a TSP tour of length at most $4n/3$.  We now show how to
apply this theorem to some special classes of graphs.

\section{Graphs Containing a Hamiltonian Path}\label{sec:hampath}

Recall that a Hamiltonian path in $G$ is a path that visits
each vertex in $V$ exactly once.  Note that the first and last vertices
on the path might not be adjacent vertices in $G$.  More generally,
$G$ might not be Hamiltonian.  In this section,
we show that for an unweighted graph $G=(V,E)$ with graph metric, if
$G$ contains a Hamiltonian path, then $G$ has a TSP tour of length at most
$4n/3$.  

\begin{theorem}\label{hampath}
Suppose $G$ contains a Hamiltonian path.  Then $G$ has a TSP tour 
of length at most $4n/3$.
\end{theorem}

\Proof
Suppose that the first and last vertices of the Hamiltonian path are
adjacent in the graph.  Then $G$ is Hamiltonian and, moreover, given the
Hamiltonian path, we can find this tour. 

If the first and last vertices of the Hamiltonian path are not
adjacent in $G$, then since $G$ is 2-vertex connected, we can use
Menger's theorem~\cite{menger1927allgemeinen,dirac1966short}, which
states that there are two vertex disjoint paths between any two
non-adjacent vertices in a 2-vertex connected graph.  Thus, we have a
simple cycle including the odd-degree nodes on the tree (the first and
last nodes in the Hamiltonian path) and the proof of the theorem
follows directly from applying Theorem \ref{cycle}.\qed

Since there are constructive proofs of Menger's Theorem, 
Theorem \ref{hampath} results in an efficient algorithm, assuming the
Hamiltonian path is given.

\section{Graphs Containing a Spanning Tree with $k$ Leaves}\label{sec:hampath2}

A Hamiltonian path can be viewed as a spanning tree with two leaves.
A natural extension is to ask what happens when a graph does not
contain a Hamiltonian path but rather a spanning tree with few
leaves.  Does it still have a short TSP tour?  Suppose $G$ has a
spanning tree with $k$ leaves.  If $G$ is well-connected, we can use a
well-known theorem of Dirac to obtain an upper bound on the length of
a TSP tour of $G$.

\begin{theorem}\label{k-leaves}
Suppose $G$ is $2(k-1)$-connected and contains a spanning tree
with $k$ leaves.  Then $G$ has a TSP tour 
of length at most $4n/3$.
\end{theorem}

\Proof A spanning tree with $k$ leaves contains at most $2(k-1)$
vertices with odd degree.  A theorem of Dirac states that if a graph
is $c$-vertex connected, then any subset $X \subseteq V$ of vertices
with $|X| \leq c$ is contained in some simple
cycle~\cite{dirac1952some, bondy1976graph}.  Thus, if $c=2(k-1)$, then
$G$ is $c$-connected by the assumption of the theorem.  Moreover, $G$ has at most
$c$ odd-degree vertices if it has $k$ leaves.  We can therefore let
$X$ be the set of odd-degree vertices
and the theorem follows directly from applying Theorem \ref{cycle}.
\qed

Finding a simple cycle containing $c$ vertices in a $c$-connected
graph can be done efficiently (see Chapter 9 in
\cite{bondy1976graph}).  Thus, Theorem \ref{k-leaves} results in an
efficient algorithm assuming the spanning with $k$ leaves is given.

More generally, Steiner cycles have been studied by the Graph Theory
community and if a set of vertices $X \subseteq V$ is contained in a
cycle, then the set $X$ is called {\em cyclable}.  This terminology
is attributed to Chvatal~\cite{chvatal1973new}.  Moreover,
cyclability of a graph $G$, i.e. $cyc(G)$, is the maximum number such
that every subset of at most $cyc(G)$ vertices is cyclable.  If a graph $G$
has a cyclable number $c = cyc(G)$ and it also contains a spanning
tree with at most $c/2 + 1$ leaves, then this spanning tree contains
at most $c$ odd-degree vertices.  Thus, it will contain a TSP tour of
length $4n/3$ via Theorem 1.  Considerable effort has been invested in
computing the cyclablity of certain graph classes.  For example, we
cite the following two theorems:

\begin{theorem}{\bf \cite{holton1982nine}}\label{nine-point}
For every 3-connected cubic graph $G$, $cyc(G) \geq 9$. 
   This bound is sharp (the Petersen graph).
\end{theorem}

\begin{theorem}{\bf \cite{aldred1999cycles}}\label{23-point}
For every 3-connected cubic planar graph $G$, $cyc(G) \geq 23$. 
   This bound is sharp.
\end{theorem}

Theorem \ref{nine-point} implies that if a 3-connected, cubic graph
$G$ contains a spanning tree with at most five leaves, then $G$ has a
TSP tour of length at most $4n/3$.  Theorem \ref{23-point} shows that
if a 3-connected, planar, cubic graph $G$ contains a spanning tree
with at most 12 leaves, then $G$ has a TSP tour of length at most
$4n/3$.  We remark that showing that a 3-connected cubic graph has a
spanning trees with at most five leaves as a means to bounding the
length of a TSP tour would only be an alternative approach, as it is
already known that a cubic graph has a TSP tour of length at most
$4n/3$~\cite{aggarwal20114, SwatiGms, boyd2011tsp, DBLP:conf/focs/MomkeS11}.

A well-known theorem of Dirac states that every graph with minimum
degree at least $n/2$ is Hamiltonian.  A analogous theorem can be
shown for cyclability.  
Let $X \subseteq V$ be a subset of vertices and define
 $\sigma_2(X) :=
\min\{\sum_{y \in Y} d(y) : Y \subseteq X, |Y|=2, Y {\text{ is an independent
set}}\}$.  
In other words, if we choose
each pair of non-adjacent vertices in $X$ and add up their degrees,
$\sigma_2(X)$ is the minimum of this quantity.  This is used in the
following theorem due to Shi:
\begin{theorem}{\bf \cite{ronghua19922}}\label{shi}
Let $G=(V,E)$ be a 2-connected graph and $X \subset V$.  If
$\sigma_2(X) \geq n$, then $X$ is cyclable in $G$.
\end{theorem}
If we find a spanning tree $T$ such that all non-adjacent pairs of
vertices with odd-degree in $T$ have total degree at least $n$ (in
$G$), then $G$ has a TSP tour of length at most $4n/3$.  The vertices
that have an even degree in the tree are allowed to have low degree in
$G$.  Another nice theorem on cyclability is due to Fournier:
\begin{theorem}{\bf \cite{fournier1985cycles}}\label{fournier}
Let $G$ be a 2-connected graph and $X \subseteq V$.  If $\alpha(X)
\leq \kappa(G)$, then $X$ is cyclable in $G$.
\end{theorem}
Here, $\alpha(X)$ means the largest independent set in $X$, and
$\kappa(G)$ is the connectivity of $G$.  It is known that if
$\alpha(G) \leq \kappa(G)$, then $G$ is
Hamiltonian~\cite{chvatal1972note}.  Theorem 
\ref{fournier} implies that if the set of odd-degree
vertices in a spanning tree has a maximum independent set that is
smaller than the connectivity of $G$, then $G$ has a TSP tour of
length at most $4n/3$.

In relation to Theorem \ref{k-leaves}, it is reasonable to ask if, for
sufficiently large $k$, a $2(k-1)$-connected graph has a spanning tree
with $k$ leaves.  This is not the case as demonstrated by the
following example.  Consider the complete bipartite graph $G =
K_{c,n}$ where $n >>c$.  Then $G$ is $c$-connected, but the minimum
length TSP tour is roughly $2n$.  So $G$ cannot contain a spanning
tree with at most $c/2+1$ leaves.

\subsection{Graphs Containing a $k$-Leaf DFS Spanning Tree}\label{sec:k-leaves}

If $G$ has a depth-first-search (DFS) spanning tree with $k$ leaves,
then we note that the techniques of M\"omke and
Svensson~\cite{DBLP:conf/focs/MomkeS11} can be used to obtain a TSP
tour of length at most $4n/3 + 2k/3$.  Specifically, in this case, it
is not difficult to see that there is a circulation (as defined by
M\"omke and Svensson) of cost at most $k$.  This implies that one can
also use the techniques from M\"omke and Svensson to prove Theorem
\ref{hampath}.  We emphasize that a DFS spanning tree must be used to
directly apply the techniques of M\"omke and Svensson.  In comparison,
in Theorem \ref{k-leaves}, we can use any spanning tree with $k$
leaves.  The proof of Lemma \ref{DFS} is straightforward, but we
include it for the sake of completeness.

\begin{lemma}\label{DFS}
If $G$ has a DFS spanning tree with at most $k$ leaves, then it has a
circulation, as defined by M\"omke and
Svensson~\cite{DBLP:conf/focs/MomkeS11}, of cost at most $k$.
\end{lemma}

\Proof We will demonstrate a 2-connected subgraph of $G$ such that the
cost of a circulation on this subgraph is at most $k$.

Consider a path from the root of the DFS tree to a leaf.  Let us call
this path $p_1$.  Suppose that the vertices on $p_1$ are labeled
sequentially from the root to the leaf in increasing order, $1,2,
... \ell(p_1)$, where $\ell(p_1)$ denotes the number of vertices in
the path $p_1$.  We find a back-edge from the leaf or the vertex
labeled $\ell(p_1)$ to a vertex with the smallest label.  Suppose that
this edge goes from $\ell(p_1)$ to $h$.  Then at the next step, we
find the back-edge $(i,j)$ where $\ell(p_1) > i > h$ and $j < i$ and
$j$ is as small as possible.  Since $G$ is 2-connected, we will always
be able to find such an edge.  Otherwise $G$ would contain a cut
vertex, which would contradict the 2-connectivity of $G$.

Now consider a path on the DFS tree from some vertex on $p_1$ to
another leaf.  Call the path from the root to this leaf $p_2$.
Perform the same procedure as above: starting at the leaf, find some
back-edges, so that the resulting subgraph containing paths $p_1$ and
$p_2$ and these back-edges is 2-connected.  At some point, we will add
a back edge that intersects with the path $p_1$.  If this is a
branching node, i.e. the last node that belongs to both $p_1$ and
$p_2$, we will add one more back edge so that the resulting subgraph
is 2-connected.

Note that each vertex in $p_2$ that is below this branching node,
i.e. has a higher label, has only one back-dge coming into it.  The
only vertices that may have more than one back-edge coming into them
are the branch node and another node with a lower label.  However,
since in Lemma 4.1 of \cite{DBLP:conf/focs/MomkeS11}, each subtree of
a branch node is accounted separately in the circulation network, if
the branch node now has, say, two back-edges, it also has two
subtrees, so its contribution to the circulation is still zero.  A
node above the branch node with $B$ back-edges coming into it will
contribute at most $B-1$ to the cost of the circulation.

As we add each root-leaf path in the DFS tree, and we add the new path
and a set of back-edges to make the subgraph 2-connected, we will add
at most one back-edge to a vertex that already has incoming
back-edges.  Thus, the circulation is upper bounded by $k$ if the DFS
tree has $k$ leaves.\qed

\begin{theorem}
If $G$ has a DFS spanning tree with at most $k$ leaves, then it has a
TSP tour of at most $4n/3 + 2k/3$.
\end{theorem}

\Proof This follows from Lemma \ref{DFS} and Lemma 4.1 of M\"omke and
Svensson~\cite{DBLP:conf/focs/MomkeS11}.  \qed

\section{Tradeoff Between Cycle Length and Approximation
  Ratio}\label{sec:general} 

We have shown that a simple cycle that contains the odd-degree nodes
in some spanning tree yields a TSP tour of length at most $4n/3$.
Suppose we can only obtain an approximate Steiner cycle.  Then what is
the guarantee on the length of the TSP tour?  We now show that we can
obtain the following tradeoff.  For a cycle $C$ in $G$ that is not
necessarily simple, recall that $|C|$ is the number of unique vertices
in the cycle $C$ and $\ell(C)$ denotes its length.

\begin{theorem}\label{tradeoff}
Given $G$, a minimum spanning tree $T$ and an approximate Steiner
cycle $C_T$ that contains all the odd-degree vertices in $T$ such that
$\ell(C_T) \leq (1 + \gamma)|C_T|$, we can construct a TSP tour of $G$
of length at most $\frac{4n}{3-\gamma}$.
\end{theorem}

\Proof We consider two cases based on the number of unique vertices in
the cycle $C_T$:
\begin{itemize}
\item[(i)] {$|C_T| > \frac{2n}{3-\gamma}$.  Then we contract the cycle $C_T$ to a
  single vertex, find a minimum spanning tree on the resulting graph
  and double each edge in this spanning tree.  Since the length of
  cycle $\ell(C_T) \leq (1+\gamma)|C_T|$, 
the total length of the
  resulting Eulerian tour is at most:
\begin{eqnarray}
\ell(TSP) & \leq & (1 + \gamma)|C_T| + 2(n - |C_T|)\\ & = & 
2n  + (1 + \gamma - 2)|C_T|\\  
& = & 2n - (1 - \gamma)|C_T|\\
& < & 2n - \frac{2n}{(3-\gamma)} (1 - \gamma)\\
& = &  \frac{4n}{3-\gamma}.
\end{eqnarray}}
\item[(ii)] $|C_T| \leq \frac{2n}{3-\gamma}$.
In this case, we find a matching of the odd-degree vertices in $T$
with length at most $(1 + \gamma)|C_T|/2$.  The total length of the resulting
Eulerian tour $S$ is at most:
\begin{eqnarray}
\ell(TSP) & \leq & n + (1 + \gamma)\frac{|C_T|}{2}\\ 
& \leq & n + \frac{(1 + \gamma)}{2}\frac{2n}{(3-\gamma)}\\
& = & \frac{4n}{3-\gamma}.
\end{eqnarray}
\end{itemize}
\qed

\subsection{Approximation Guarantees from LP Bounds}

In general, it could be the case that there does not exist a spanning
tree whose odd-degree vertices can be contained in a simple cycle.  An
example of such a graph can be found in Figure \ref{fig:general}.
\begin{figure}[h]
\begin{center}
\epsfig{file=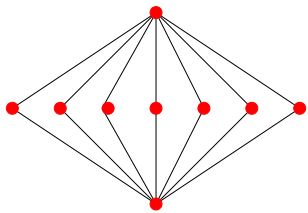, width=4cm} 
  {\caption{Any spanning tree of this graph has too many leaves to be
      spanned by a simple cycle.  However, note that the solution to
      the Held-Karp LP relaxation will be $|E| = 2n$ for this graph,
      certifying that the lower bound is much greater than $4n/3$ in
      this case.}\label{fig:general}}
\end{center}
\end{figure}
However, suppose we can compute, via an LP relaxation or some other means, a
lower bound on the length of a TSP tour, e.g. $ OPT \geq (1+ \alpha)n$
for $0 \leq \alpha \leq 1$.  Then the following Corollary of Theorem
\ref{tradeoff} states a sufficient condition for a
$\frac{4}{3}$-approximation to the optimal TSP tour.

\begin{corollary}\label{tradeoff1}
If an optimal tour is lowerbounded by $OPT \geq (1+ \alpha)n$ and $G$
contains a spanning tree $T$ and a cycle $C_T$ containing the
odd-degree nodes of $T$ such that $\ell(C_T) \leq (1+4\alpha)|C_T|/(1+
\alpha)$, then $G$ has a TSP tour of length at most $\frac{4}{3} \cdot
OPT$.
\end{corollary}

Note that Theorem \ref{tradeoff} says that if we can find a tree $T$
and a cycle $C_T$ such that $\ell(C_T)/|C_T| < 4/3$, then we can find
a TSP tour less than $3n/2$.  To find a tour shorter than $7n/5$
(which is currently the best known bound when the solution to the
standard LP relaxation equals $n$~\cite{sebHo2012shorter}), we require
that $\ell(C_T)/|C_T| < 8/7$.

\section{Discussion}\label{sec:discussion}

We have reduced the problem of finding a short TSP tour to the problem
of finding an (approximate) Steiner cycle where the required vertices
are the odd-degree nodes in some spanning tree, and we have
flexibility as to whether or not we include the non-required vertices
in the cycle.  But is this problem any easier than graph-TSP itself?
For example, in Figure \ref{tsp-mst}, we give an example of a graph
and a spanning tree such that the odd-degree vertices of the spanning
tree is the entire vertex set!  Thus, finding a Steiner cycle for
these vertices is no easier than finding a TSP tour.

\begin{figure}[h!]
\begin{center}
\epsfig{file=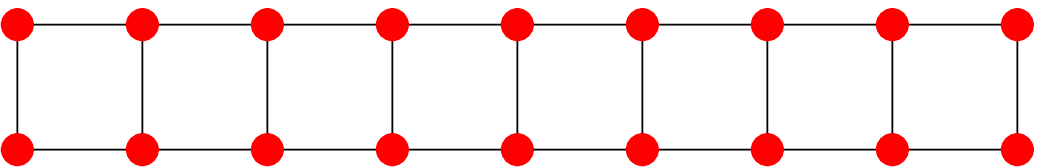, width=5cm} 
\hspace{10mm}
\epsfig{file=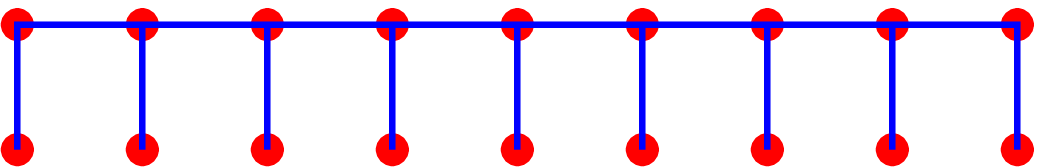, width=5cm}
\end{center}
\caption{A graph $G$ and a spanning tree.}\label{tsp-mst}
\end{figure}
However, in this example, we can see that there are many other possible
spanning trees.  Figure \ref{tsp-mst2} shows two other possible
spanning trees and corresponding
Steiner cycles.  
\begin{figure}[h!]
\begin{center}
\epsfig{file=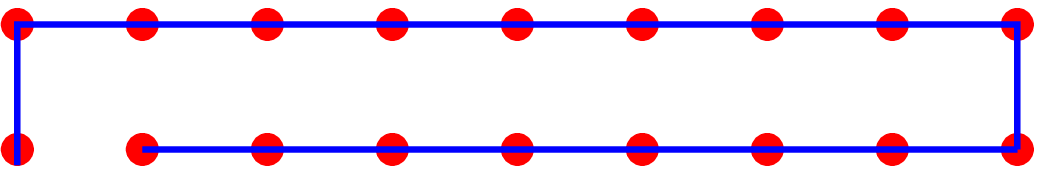, width=5cm} \hspace{10mm}
\epsfig{file=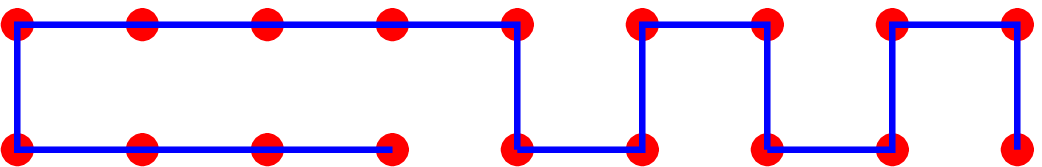, width=5cm} \\
\vspace{5mm}
\epsfig{file=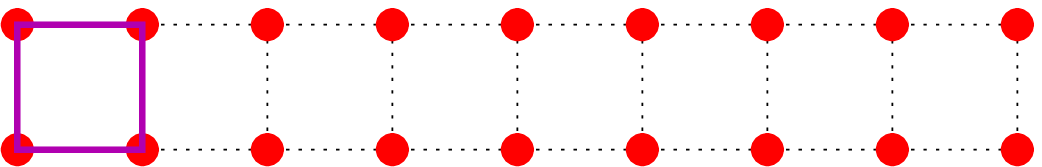, width=5cm} \hspace{10mm}
\epsfig{file=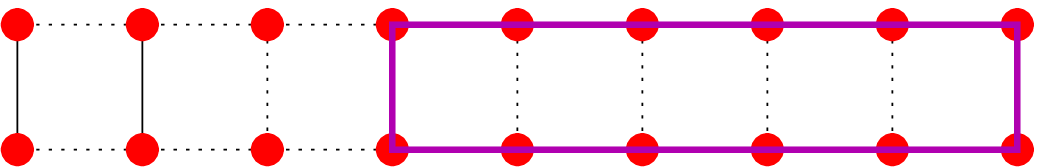, width=5cm}
\end{center}
\caption{Alternative spanning trees for $G$ and corresponding Steiner
  cycles.}\label{tsp-mst2}
\end{figure}
We note that given a spanning tree, the Steiner cycle including the
odd-degree nodes may not be unique.  Another example of a graph $G$
and a spanning tree in which every vertex can have odd degree is shown
in Figure \ref{fig:wheel}.  But, again, there are many other spanning
trees in which only a subset of the vertices have odd degree.
\begin{figure}[h!]
\begin{center}
\epsfig{file=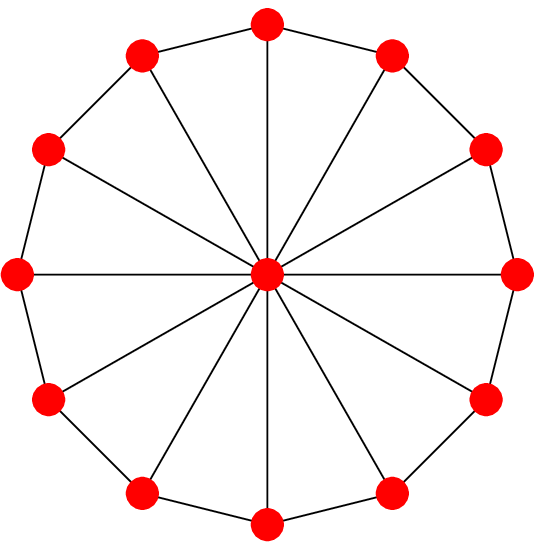, width=3cm}
\hspace{5mm} 
\epsfig{file=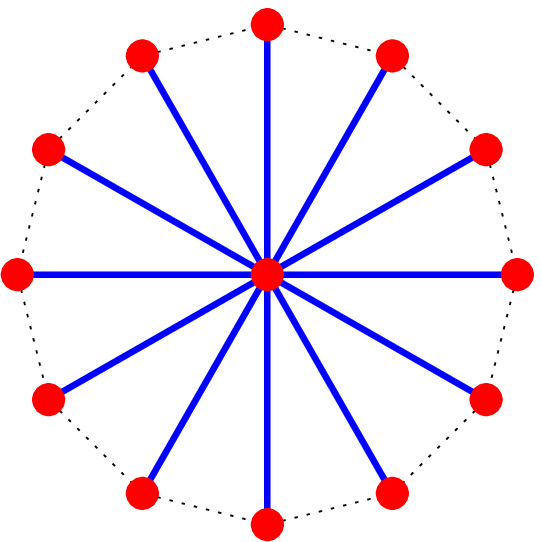, width=3cm} 
\end{center}
\caption{The wheel graph has a spanning tree in which all vertices
  have odd degree.}\label{fig:wheel}
\end{figure}
\begin{figure}[h!]
\begin{center}
\epsfig{file=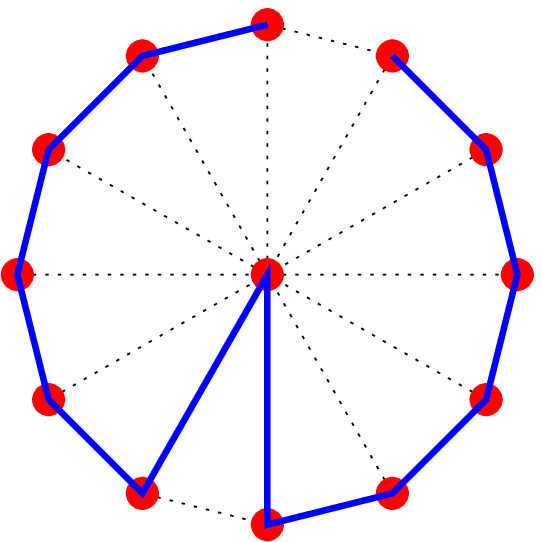, width=3cm}
\hspace{5mm}
\epsfig{file=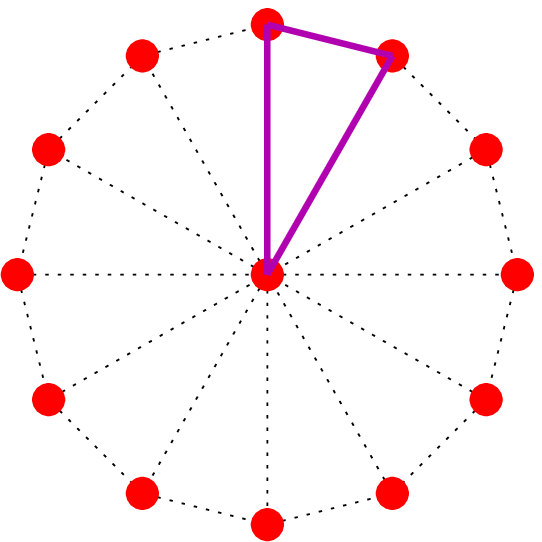, width=3cm}
\hspace{5mm}
\epsfig{file=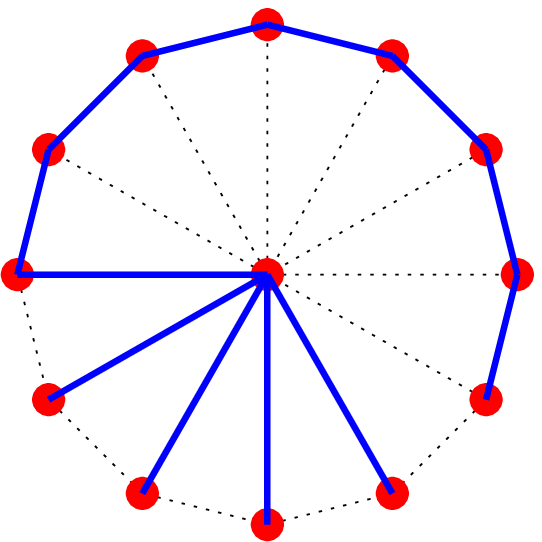, width=3cm}
\hspace{5mm}
\epsfig{file=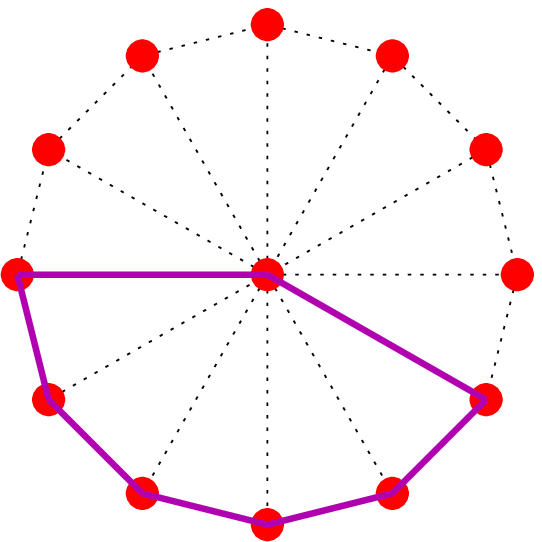, width=3cm}
\end{center}
\caption{Alternative spanning trees with fewer odd-degree vertices for the wheel graph.}\label{fig:wheel2}
\end{figure}

Each of the examples we have considered so far actually contains a
Hamiltonian path.  Thus, by applying Theorem \ref{hampath}, we can see
that they have a TSP tour of length at most $4n/3$.  There are
actually interesting examples of cubic, 3-edge connected graphs that
do not contain a Hamiltonian path.  The graph shown in Figure
\ref{fig:zam} is such a graph due to
Zamfirescu~\cite{zamfirescu1980three}.  We see that we can construct a
spanning tree and Steiner cycle containing all of the vertices that
have odd degree in the spanning tree.
\begin{figure}[h!]
\begin{center}
\epsfig{file=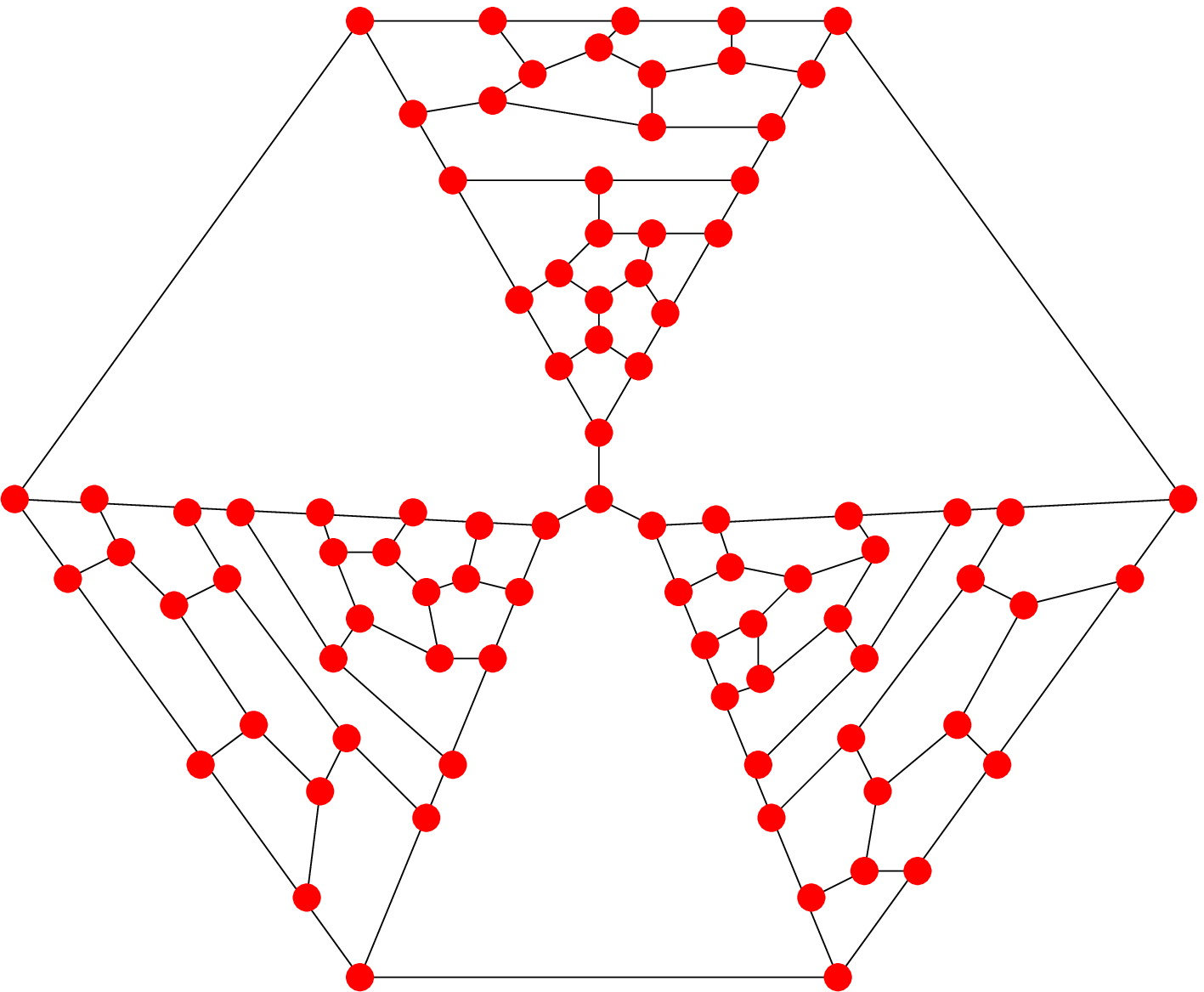, width=5cm} \hspace{5mm}
\epsfig{file=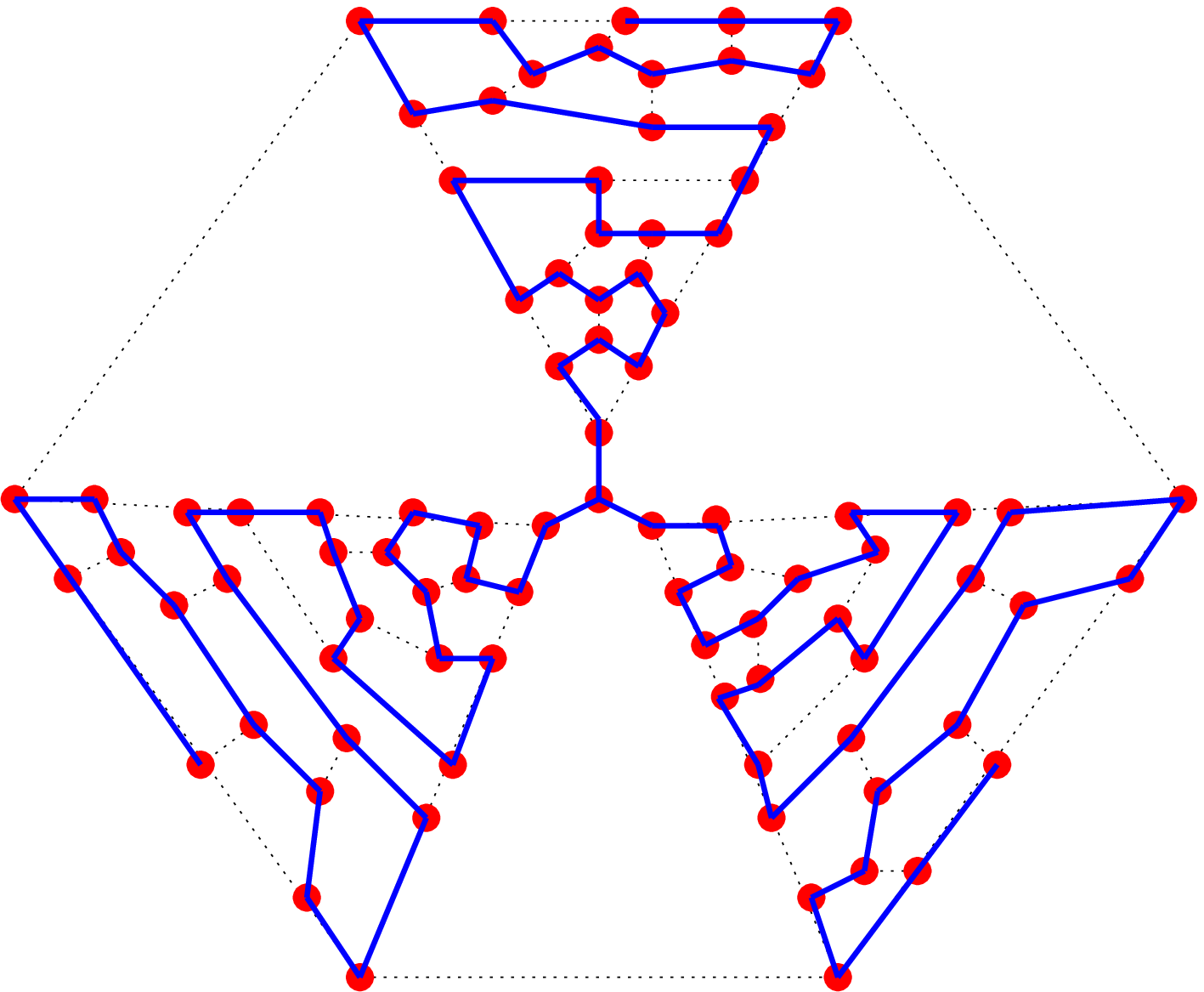, width=5cm} \hspace{2mm}
\epsfig{file=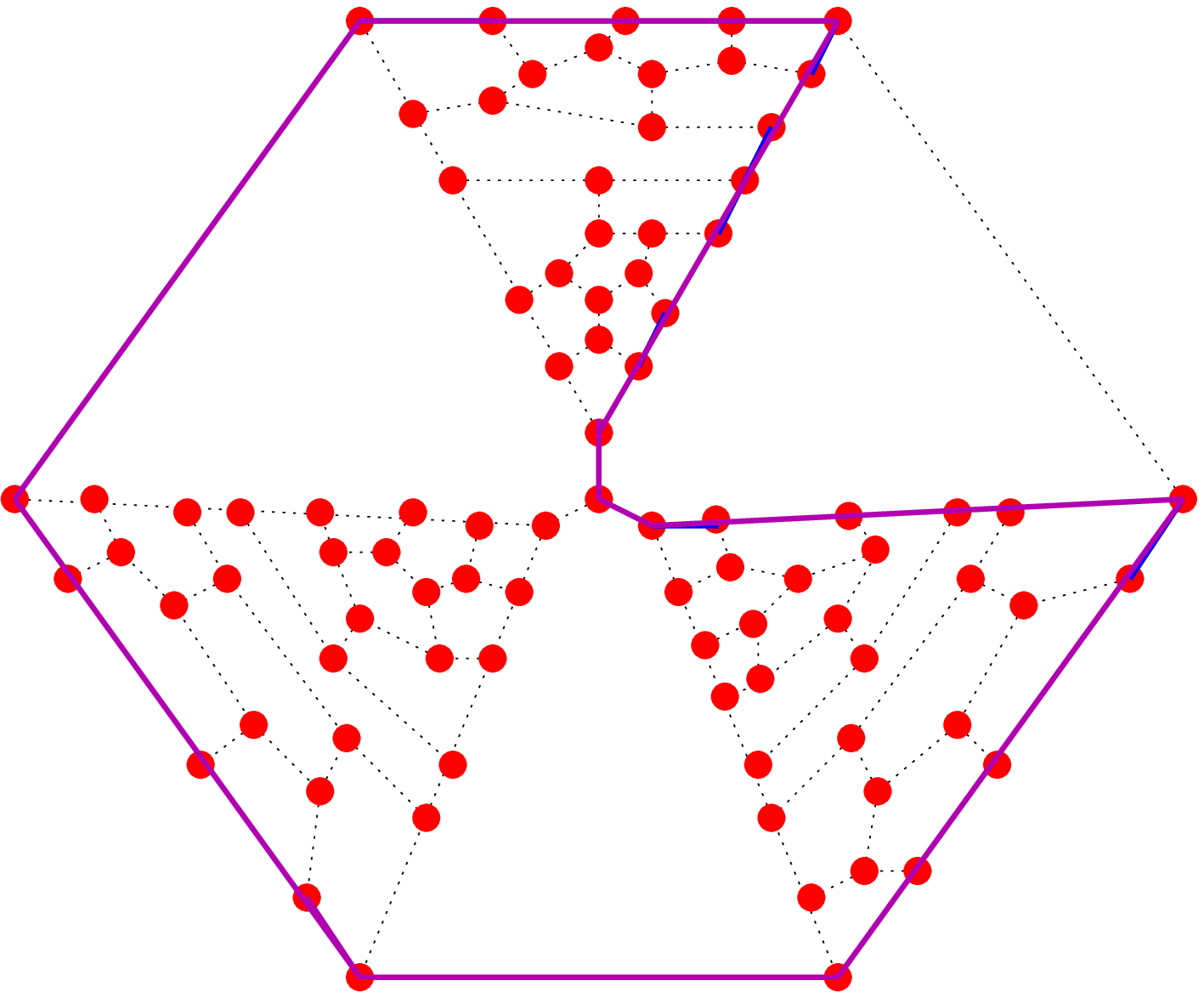, width=5cm} 
\end{center}
\caption{A cubic, 3-edge connected graph with no Hamiltonian paths.
  We show a spanning tree and a corresponding Steiner cycle containing
  all the nodes with odd degree in the spanning tree.}\label{fig:zam}
\end{figure}

In conclusion, let us consider the following question: Suppose the
standard linear programming relaxation for Graph TSP has value $n$ on
a fixed graph.  Then is there a spanning tree $T$ and a simple cycle
$C_T$ that contains all of the vertices that are odd-degree in $T$?
If a graph is Hamiltonian, then this is (trivially) true for any
spanning tree.

\section*{Acknowledgements}
We would like to thank the anonymous referees for many useful comments.

\begin{small}
\bibliography{steiner-cycle}
\end{small}
\end{document}